# Light pollution in shallow coastal waters: aquaculture farming in the Galician Atlantic shoreline

**Salvador Bará,**[1,*] **M. Luz Pérez-Parallé,**[2] **José L. Sánchez,**[2] **Raul Cerveira Lima,**[3,4,5] **Xurxo Cegarra,**[6] **and José Antonio Fernández Bouzas**[7]

[1] *Light Pollution Lab, Dept. Física Aplicada, Universidade de Santiago de Compostela. 15782 Compostela, Galicia.*

[2] *Laboratorio de Biología Molecular y del Desarrollo. Departamento de Bioquímica y Biología Molecular. Instituto de Acuicultura. Universidade de Santiago de Compostela. 15782 Compostela, Galicia, Spain.*

[3] *Physics, Escola Superior de Saúde, Politécnico do Porto, Portugal.*

[4] *CITEUC – Centre for Earth and Space Research, University of Coimbra, Portugal..*

[5] *CISA – Centro de Investigação em Saúde e Ambiente, Politécnico do Porto, Portugal.*

[6] *Wireless Galicia, SL. Rúa Colon 26,  4, Of. 3.  36201 Vigo, Galicia,Spain.*

[7] *Director Conservador Parque Nacional das Illas Atlánticas de Galicia. Consellería de Medio Ambiente e Ordenación do Territorio. Rúa Oliva 3. 36202 Vigo, Galicia, Spain.*

[*]*Corresponding author: salva.bara@usc.es*

**Abstract**

Artificial light at night is a relevant environmental stressor whose unwanted consequences have been documented by a growing body of research. In this work we analize the loss of the natural darkness of the night in aquaculture farms located in shallow waters close to the shoreline, in the Atlantic coast of Galicia. High-luminance LED lampposts are switched-on the whole night long for surveillance purposes, significantly increasing the light levels throughout the whole water column, from the surface to the seabed. Measured in terms of the weighted radiance associated with the significant magnitude $m_{1/3}$, the zenithal brightness at these locations is 125 times higher than should correspond to a pristine natural site. Given the potential severity of the environmental effects of artificial light at night, the dubious efficacy of light *per se* as a theft deterrence factor, and the availability of more efficient and environmentally friendly technologies for surveillance, the use of permanent floodlight sources in coastal areas is strongly discouraged and should be restricted as much as possible.

# 1. Introduction

Artificial lighting is a key factor for societal well-being and economic development. It has been instrumental for enabling the expansion of human activities into the nighttime, relieving us from the dependence on the natural cycles of light imposed by the relative movements of the Sun-Earth-Moon system. However, despite its clear advantages, the use of light is not free of unwanted and relevant side effects. In the last century, and more intensely in the last decades, the pervasive growth of the anthropogenic light emissions (Cinzano et al., 2001; Falchi et al., 2016; Kyba et al., 2015) has given rise to a substantial increase of the illumination levels in many natural areas that should remain dark (Bará, 2016; Davies et al., 2016; Jechow et al., 2016). Key aspects of life on the surface of Earth, evolutionarily adapted to these natural light cycles, can be severely disrupted by the intrusion of artificial light during the twilight and the night periods (Foster et al., 2018; Hölker et al., 2010a, 2010b; Gaston et al., 2013, 2014; Longcore and Rich, 2004; Navara and Nelson, 2007; Rich and Longcore, 2006; Sanders and Gaston, 2018). Artificial light at nighttime acts in many aspects as a conventional polluting agent, and there is a growing consensus that it should be handled accordingly.

Coastal regions are particularly affected by this emerging environmental stressor (Bolton et al., 2017; Davies et al. 2014, 2015, 2016; Ges et al., 2018; Jechow et al., 2017; Rodríguez et al., 2017; Tamir et al., 2017). They are home to a rich biodiversity, with a wealth of interrelated light-sensitive processes taking place at the land-water boundary. Coastal regions, besides, tend to be densely populated in many places of the world, harbouring a diverse mix of economic activities that include fishing, aquaculture, maritime transportation, tourism destinations and shipyard industry, among others. All these activities pose additional lighting needs, well beyond those required by the normal functioning of the urban settlements near the shoreline. As a result, the basic day-night photic cycle can be severely distorted during nighttime, and the monthly cycle of light associated with the phases of the Moon, which is also ecologically relevant (see, e.g. Fallows et al., 2016; Kyba et al., 2017a; Rivas et al., 2015; Zantke et al., 2013), may be blurred to the point of becoming almost unrecognizable (Bará, 2016; Posch et al., 2018; Davies et al., 2013).

A particularly challenging situation arises in shallow-water open-farm bivalve mollusk aquaculture, an essential economic asset for many coastal communities. In this kind of farms, located in the intertidal and shallow subtidal zones, artificial light at night is used for a variety of purposes, including enhanced visual performance for work and safety, detection of unauthorized access, and (allegedly) theft deterrence. The availability of new and efficient solid-state lighting sources (LED) and the perceived low-cost of the light produced by them has given rise to a sustained increase of lighting levels and to the expansion of the illuminated areas, a trend also observed at global scale (Kyba et al. 2017b). Not infrequently, powerful and intense security lamps are lit all night long, bathing with a flood of light the water column from the surface to the seabed. Light pollution has been reported to produce significant effects on relevant aspects of the physiology and behavior of freshwater and marine species (Brünig et al., 2017, 2018; Castorani et al., 2015; Dimitriadis et al., 2018; Luarte et al., 2016; Ludvigsen et al., 2018; Manfrin et al., 2018; Moore et al., 2000; Nielsen and Strömgren, 1985; Pendoley and Kamrowski, 2015, 2016; Robson et al., 2010; Strömgren, 1976a, 1976b; Thums et al., 2016; Tuxbury and Salmon, 2005; Underwood et al., 2017; Waissel et al., 1999). Additional concerns arise from the potentially detrimental effects associated with the changes in the spectral composition of light: the replacement of the traditional low correlated color temperature (CCT) gas-discharge sources (e.g. high-pressure sodium vapor, HPS) with high CCT LED lamps involves introducing substantially higher amounts of power in the short-wavelength (blue) region of the optical spectrum (Aubé et al., 2013). Light in this spectral region is efficiently transmitted throughout the water column, and many taxa are particularly sensitive to it.

We address in this paper the loss of the natural levels of darkness set off by the use of security lights in this kind of farms. As a significant case-study, we analyze a relevant aquaculture area located at Carril, within the Ría de Arousa bay, in the Galician Atlantic coast. The cultivation area under the influence of the lighting system above mentioned comprises an intertidal coastal zone of approximately 1 million m². It is the only coastal area in the northwest of Spain where there are private concessions for the cultivation of bivalve mollusks. There are references about the existence of mollusk culture in this area during the last 500 years. 656 private farmers cultivating 1283 plots are dedicated mainly to the exploitation of different species of clams and cockles. According to Xunta de Galicia (2017), 650 tons of the carpet shell clam *Ruditapes phillipinarum* [Adams and Reeve, 1850], 140 tons of the pullet carpet clam *Venerupis corrugata* [Gmelin, 1791] and 45 tons of the grooved carpet shell *Ruditapes decussatus* [Linnaeus, 1758] were produced in 2017. In addition, the production of the cockle *Cerastoderma edule* [Linnaeus, 1758] was approximately of 250 tons. Overall this artisanal production represents an income of 10-15 million euros and it also has socio-economic importance because it involves numerous fishermen and small family businesses. It is important to note that the special environmental conditions of the area are optimal for the cultivation of these bivalve mollusks, thus the meat yield obtained is usually 30% higher than other regions of Spain or Europe. The natural restoration of the parks traditionally means 30-40% of the total production which is completed with seed from hatcheries. In this site, the overall light levels above the surface and on the seabed were continuously monitored during a campaign that spanned several weeks in the summer of 2017, and compared against the values expected for a pristine natural night. Reference night brightness levels for natural sites and urban areas in the same nights of the measurement campaign were provided by the sensors of the *Galician Night Sky Brightness Monitoring Network*, jointly operated by MeteoGalicia, the public meteorological agency of the Galician government (Xunta de Galicia), and the Light Pollution Laboratory of the Universidade de Santiago de Compostela (Bará, 2016; MeteoGalicia, 2018).

The structure of this paper is as follows: in section 2 we describe the measurement site and the main characteristics of the radiance meters used in the campaign. The results are described in section 3, using a variety of graphical formats for an easier visualization of the disruption of the natural night darkness at that site. Discussion and conclusions are drawn in sections 4 and 5, respectively.

## 2. Materials and Methods

*2.1. Measurement site*

This study was carried out in a large aquaculture area located at 42.6081° N, 8.77735° W, in Carril, close to the coastal town of Vilagarcia de Arousa, Galicia, NW of Iberian Peninsula (Fig.1). This bivalve culture area comprises both shallow waters with typical depths 1-2 m (relative to the datum defined by the local lowest astronomical tide, LAT), and extensive sand shoals located 0.7-1 m above LAT. Tide amplitudes easily reach several meters, so the whole zone, including the shoals, is most of time underwater. The cultivation of clam and cockle species, mentioned above, is carried out extensively and with frequent addition of seed from hatcheries. The density of seed ranges between 2 and 5 kg/m$^2$ depending on the size of the mollusks. The average production is between 10-12 kg/m$^2$ at the time of extraction. The producers are responsible for the cleaning, maintenance and seeding of the parks, as well as the control of predators.

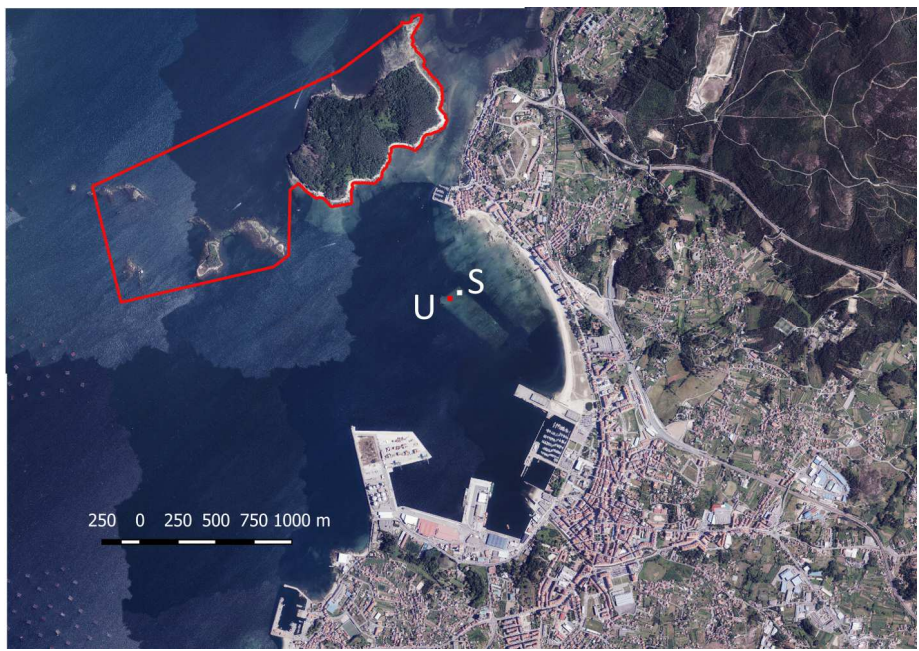

**Fig. 1.** The towns of Carril (top) and Vilagarcía de Arousa (bottom), in the inner south-eastern region of Ría de Arousa, Galicia. The dots show the locations of the underwater (U) and surface (S) radiance measurement sensors. The area enclosed within the full line corresponds to the arquipelago of Cortegada, a set of small islands belonging to the Galician Atlantic Islands

Maritime-Terrestrial National Park. Orthographic PNOA image © Instituto Geográfico Nacional - Xunta de Galicia.

The farming area is illuminated by 23 dedicated lampposts, evenly distributed along the shoreline and across the sea. Every lamppost is fitted with several high-luminance LED panels (typically composed of 150 individual LED emitters each) that are switched on all night long for surveillance purposes. During normal operation the whole water column, from the surface to the seabed, is brightly lit (Fig.2).

Two radiance meters (see details in section 2.2., below) were installed in this farm during the months of July and August, 2017. The selected place was a shoal whose floor lies 0.7 m above LAT, 560 m distant from the shoreline (Fig. 1). The underwater detector, U, was located close to the bottom, 0.6 m above the seabed (1.3 m above LAT), whereas the surface detector, S, was installed on the top of a pole, 3 m above the seabed (3.7 m above LAT). The nearest LED lamppost was 35 m away from the U detector, and 80 m distant from the S one.

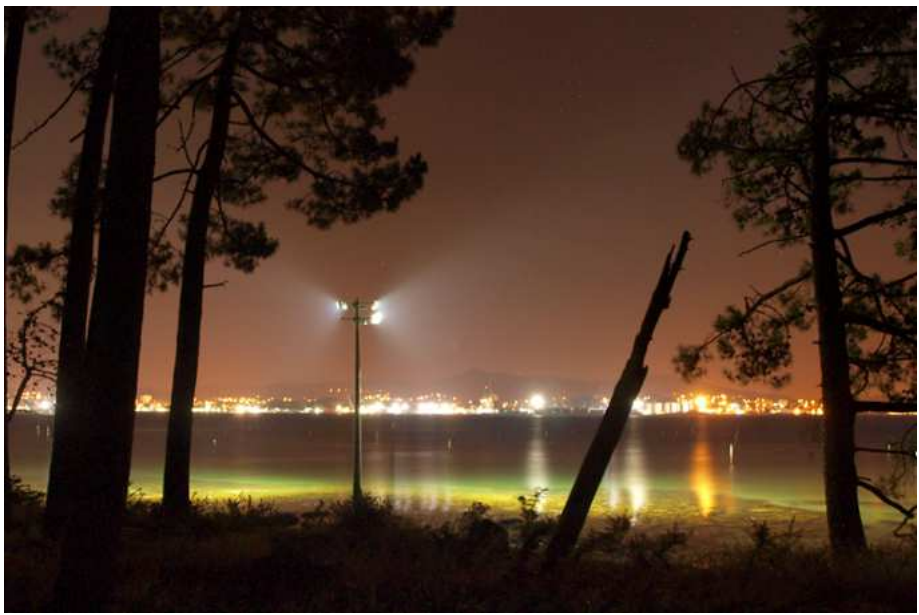

**Fig. 2**. Typical illumination pattern of high-luminance LED surveillance lampposts (30 s exposure at f/5.6, f = 14mm, ISO 400), as seen from Cortegada island.

The night sky brightness changes from day to day, and continuously within each night, due to the variable contributions of the artificial light sources (streetlights, residential, advertisement, industrial, and vehicle lights) and the natural ones (celestial objects above the observer, natural airglow), mediated through the ever changing state of the atmosphere (presence or absence of different types of clouds, and varying aerosol concentration profiles, which determine the absorption and scattering of the light beams). For a better appraisal of the loss of the natural night conditions at the Carril location, the data obtained by the underwater and surface sensors are compared here against the data recorded during the very same nights by identical sensors installed in MeteoGalicia weather stations at three other places, representative of increasing levels of light pollution due to artificial light scattered in the atmosphere: Xares (42.2078° N, 6.89256° W), an inland site in the Eastern Galician mountains with very low levels of light pollution and near-pristine natural dark skies; the Cíes Islands (42.2118° N, 8.90842° W), located at the entrance of the Ría de Vigo, belonging to the Galician Atlantic Islands Maritime-Terrestrial National Park; and the city of Vigo (42.2417° N, 8.72759° W), a heavily light polluted and densely populated metropolitan area in the southern coast of Galicia. The joint analysis of the data recorded by all these sensors provides interesting insights to assess the degradation of the natural night darkness at the Carril site.

*2.2. Radiance meters*

We used the zenithal night sky brightness (NSB) as a proxy variable for characterizing the anthropogenic disruption of the natural night at the observing site. "Brightness", in this context, is a short-hand term for the spectral radiance of the night sky, spatially averaged within the field of view of the detector and spectrally integrated within its photometric band. The zenithal NSB measurements were made using low-cost SQM-LU-DL light meters (Unihedron; Ontario, Canada), a radiance sensor with data-logging capability, based on a TSL237 high-sensitivity irradiance-to-frequency converter chip with temperature correction (TAOS, USA). This detector is fitted with optics restricting its field of view to a region of the sky with approximately Gaussian profile and full-width-at-half-maximum (FWHM) 20°. The SQM photometric band spans the spectral interval 400-650 nm (Cinzano, 2005; Pravettoni et al., 2016). A water-tight specific enclosure for SQM-LU-DL underwater operations was designed and manufactured by Galventus, S.L. (Moaña, Galicia) (Figs. 3-4).

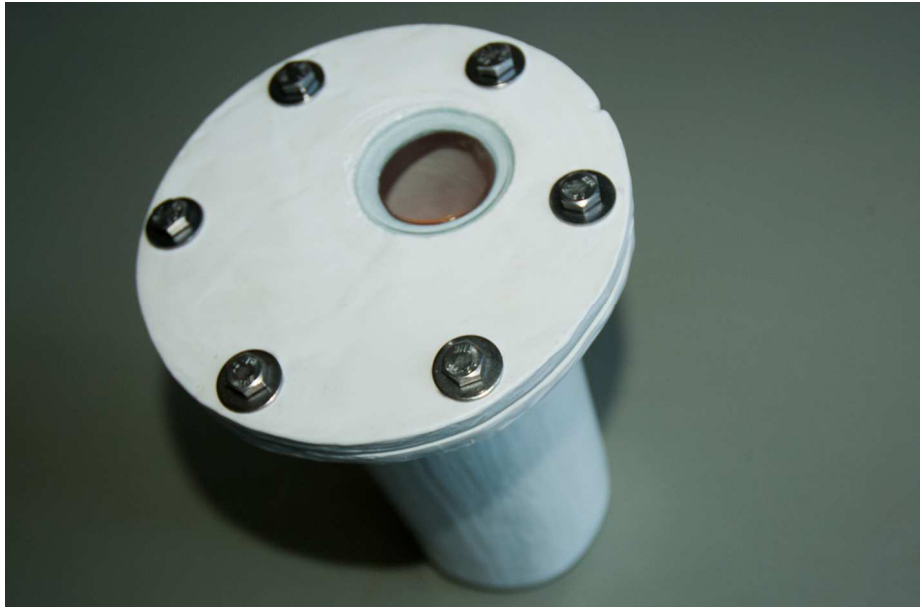

**Fig. 3**. Water-tight enclosure for underwater SQM-LU-DL radiance meters.

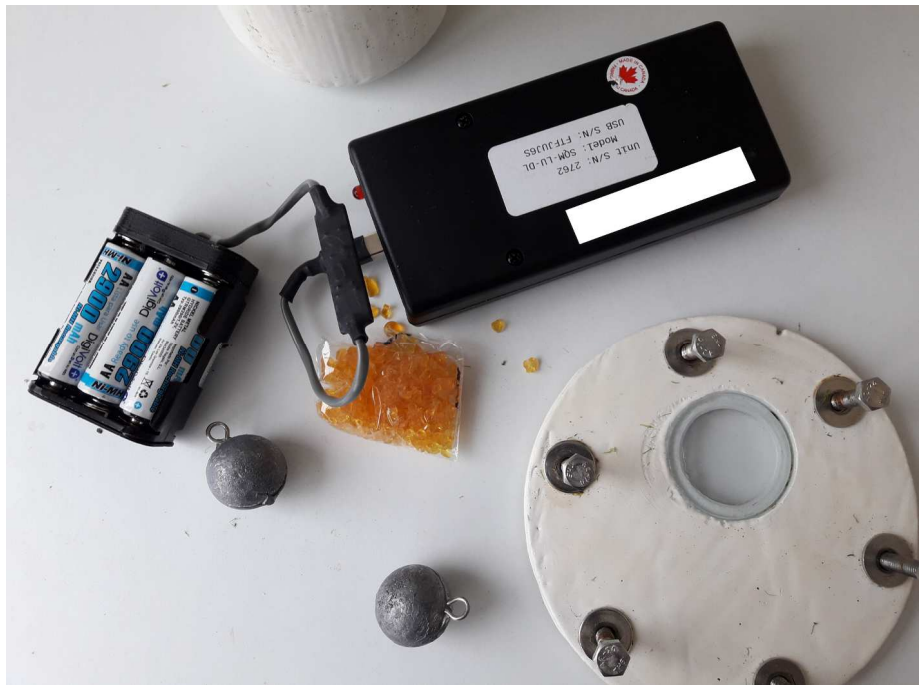

**Fig. 4**. Retrieved underwater SQM-LU-DL radiance meter, with batteries, desiccant bag, and lead ballasts samples.

The SI units for brightness are those of radiance [Wm$^{-2}$sr$^{-1}$], spectrally weighted by the sensitivity function of the photometric band in which the observations are carried out. Several photometric bands are commonly used for reporting the NSB, among them the Johnson-Cousins *B*, *R*, and *V* (Bessell, 1979), and the spectral response of the human visual system described by the CIE photopic spectral efficiency function V(λ). In this last case the spectrally-weighted radiance, after being multiplied by the scaling factor 683 lm/W (lm: *lumen*, SI unit of photopic light flux), gives us the luminance, in SI units cd/m$^2$ (cd: *candela*, fundamental SI unit for light intensity), which approximately correlates with the brightness perceived by an average human observer under foveal fixation (CIE, 1990).

The NSB is also frequently expressed in m*agnitudes per square arcsecond* (mag/arcsec$^2$), a negative logarithmic scale with longstanding tradition in Astronomy. The relationship between the SI brightness $L_V$ in a given photometric band, *V*, and the corresponding mag$_V$/arcsec$^2$, $m_V$, is given by

$$L_V = C \times 10^{(-0.4 m_V)} \qquad (1)$$

where $C$ is a zero-point constant that settles the origin of the magnitude scale (i.e., the brightness corresponding to 0.00 mag/arcsec$^2$). Note that smaller values of $L_V$ (darker skies) correspond to larger values of $m_V$ in mag/arcsec$^2$. For the human visual system the relationship between luminance and magnitudes per square arcsecond in the CIE V(λ) band, using the absolute AB magnitude scale, is given by (Bará, 2017)

$$L[cd/m^2] = 10.96 \times 10^4 \times 10^{(-0.4 m)} \qquad (2)$$

The measurements of the SQM-LU-DL sensor are given in mag$_{SQM}$/arcsec$^2$. This device has a specific photometric band, but its readings are internally corrected to approximate the magnitude values that would be recorded by a Johnson-Cousins V detector. The V band, in turn, although not strictly equal to the CIE V(λ) band, is reasonably close to it. Hence, Eq.(2) can be used to estimate the luminance values, in cd/m$^2$, corresponding to the mag$_{SQM}$/arcsec$^2$ readings provided by the sensor. It must be kept in mind, however, that this last conversion is only approximate (Bará, 2017; Sánchez de Miguel et al., 2017).

A stated above, the field of view (FOV) of the SQM-LU-DL detector has a Gaussian profile with FWHM of 20°, and hence its readings are a weighted solid-angle average of the incident radiance. Note that off-axis light sources located at angles larger than 20° from the center of the FOV, if sufficient intense, may contribute non-negligibly to the measured values. When the device is submerged the 20° FWHM in water corresponds to an effective larger FOV above the surface, due to the refraction of the light rays upon trespassing the air-water interface. If the sea surface is calm and flat, the whole 180° celestial hemisphere is shrunk underwater to a cone of directions of absolute full-width 97.2°; the effective FOV of the SQM loses its Gaussian profile and the 20° FWHM angle underwater becomes 26.8° above the surface. Note also that the waves and ripples on the sea surface act as curved refractive boundaries, so that intense light sources of small angular extent will give rise to a rich and ever-changing caustic pattern particularly noticeable on shallow water seabeds. The presence of waves can, that way, induce a large variability in the underwater recorded brightness, when compared to the one recorded above the surface under the same lighting conditions.

## 3. Results

We analyze in this section the data acquired by the surface and underwater sensors at the Carril farm site during the measurement campaign carried out from July 12th, 2017 (12:00 UT) until August 23rd, 2017 (11:59 UT). Zenithal brightness readings were continuously taken at a rate of one every three minutes and stored in the devices dataloggers. The recorded datasets were downloaded and processed after the sensors were retrieved at the end of the campaign. Both sensors were fully functional during the whole measurement period, and no data losses were experienced. Datasets coincident in time from the Xares, Ilas Cíes, and Vigo sensors of the Galician Night Sky Brightness Monitoring Network were provided by MeteoGalicia for this study. The brightness data referred to in this section correspond in all cases to the radiance at the entrance of the glass window of the protective housing of each sensor, obtained by correcting the recorded raw data for the light losses caused by the Fresnel reflections at the glass surfaces.

*3.1 NSB histograms*

Fig. 5 shows the overall histograms of the zenithal night sky brightness recorded at the five locations included in this study. The horizontal axis of each histogram corresponds to the NSB values, binned in 0.05 $mag_{SQM}/arcsec^2$ intervals, and the vertical one displays the observed relative frequency (%) of each magnitude bin. Only NSB values larger than 13.5 $mag_{SQM}/arcsec^2$, belonging to twilight and nighttime periods, are displayed, since natural daytime brightness levels due to solar radiation are not included in this study.

The histograms show several conspicuous features. All of them have a well-defined peak in the rightmost region of higher $mag_{SQM}/arcsec^2$ values, corresponding to the darkest nights experienced at each site. The mode of this peak for the mountain site of Xares is 21.6 $mag_{SQM}/arcsec^2$, revealing a well-preserved natural sky. For comparison, astronomical dark-sky reserves promoted by several international organizations shall have typical NSB values of at least 21.4 $mag_{SQM}/arcsec^2$ to be certified as such, and the darkest expected natural skies have NSB of order 22.0 $mag_{SQM}/arcsec^2$ (Falchi et al., 2016). The Illas Cíes site (21.1 $mag_{SQM}/arcsec^2$) is slightly brighter than Xares, as expected given its relative proximity to the brightly lit metropolitan area of Vigo (18.6 $mag_{SQM}/arcsec^2$, measured at a place close to the city center). The brightness levels recorded at the Carril aquaculture area are however surprisingly high: the mode of the darkest peak of the surface detector is 18.25 $mag_{SQM}/arcsec^2$, a value that would be characteristic of a more densely populated area (compare to e.g. Vigo, above), and that of the underwater one, located 35 m away from the nearest high-brightness LED surveillance lamppost, is 16.45 $mag_{SQM}/arcsec^2$. This means (see Eq. 1) that the most frequent brightness recorded at the detector located at the seabed is more than 100 times larger than the one that would be expected for a natural night in a well-preserved site above the surface (e.g., Xares).

In urban settings and other light polluted places, the darkest values of the night sky are commonly attained in clear and moonless nights. Under overcast skies the brightness at these places becomes significantly larger, due to the downward scattering of the artificial light at the base of the clouds (Kyba et al, 2012; Puschnig et al., 2014a, 2014b; Aubé et al, 2016; Bará, 2016; Solano-Lamphar and Kocifaj, 2016; Ribas, 2017). This is why the histograms of Illas Cíes, Vigo, and Carril (surface) show a clear bimodal structure, with a secondary peak associated with cloudy nights located about 3 mag$_{SQM}$/arcsec$^2$ below the first one. Again according to Eq. 1, this indicates that in overcast nights the sky at these locations is almost 16 times brighter than in clear nights. Note that the underwater sensor located at the seabed recorded very high light levels almost continuously, due to the direct light received from the surveillance lampposts. The bimodal histogram collapses in this case to a single, unimodal one, suggesting that the direct radiance from the lamps is the main contributor to the recorded brightness, and that the variations of the sky brightness, either in clear or overcast nights, play a relatively minor role to determine the light levels at the seabed.

In pristine dark places, on the contrary, moonless overcast nights turn out to be darker than moonless clear nights, since the cloud cover, not illuminated by artificial sources, acts as a screen blocking the light from celestial bodies and upper-atmosphere airglow that would otherwise reach the detector (Ribas et al, 2016).

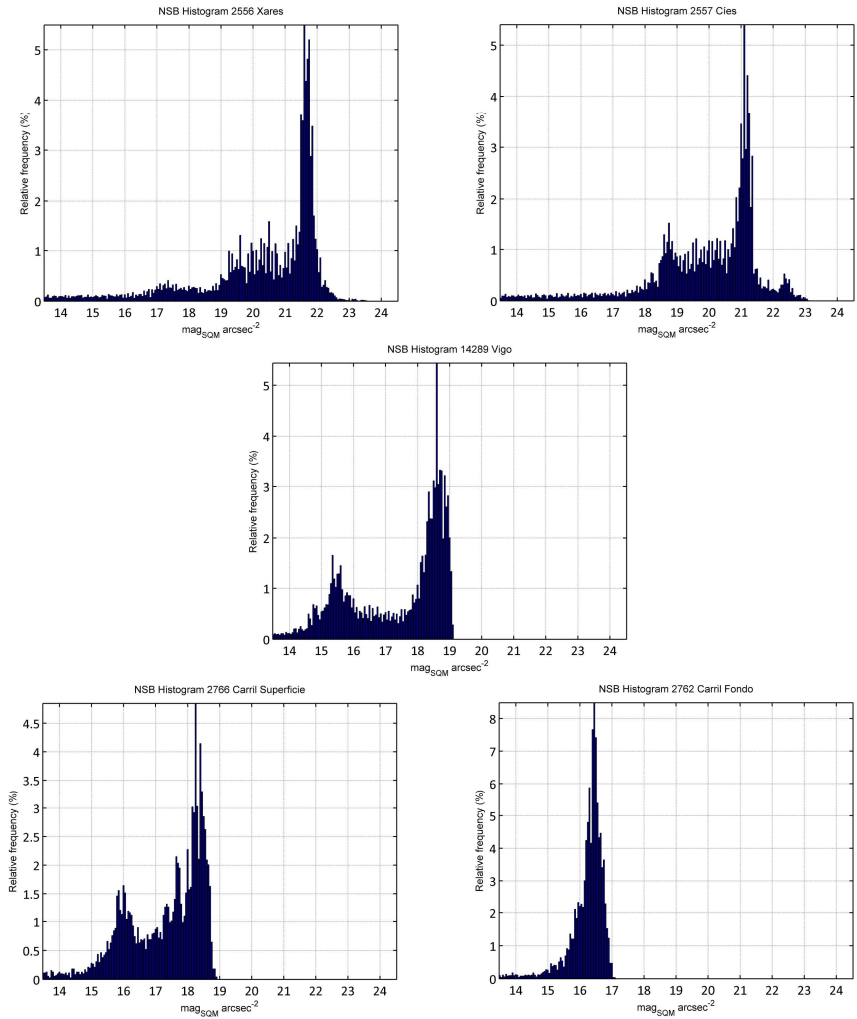

**Fig. 5**. Histograms of the recorded zenithal night sky brightness values at the stations of Xares (upper left), Illas Cíes (upper right), Vigo (center), Carril surface (lower left), and Carril seabed (lower right).

The cumulative sums associated with these histograms can be seen in Fig. 6. The values of each curve can be interpreted as the probability of observing a zenithal night sky brighter (i.e., with smaller $mag_{SQM}/arcsec^2$) than the value indicated in the horizontal axis. The behaviour of the Carril surface signal closely resembles that of a highly light polluted metropolitan area. The underwater brightness, as indicated above, is even larger.

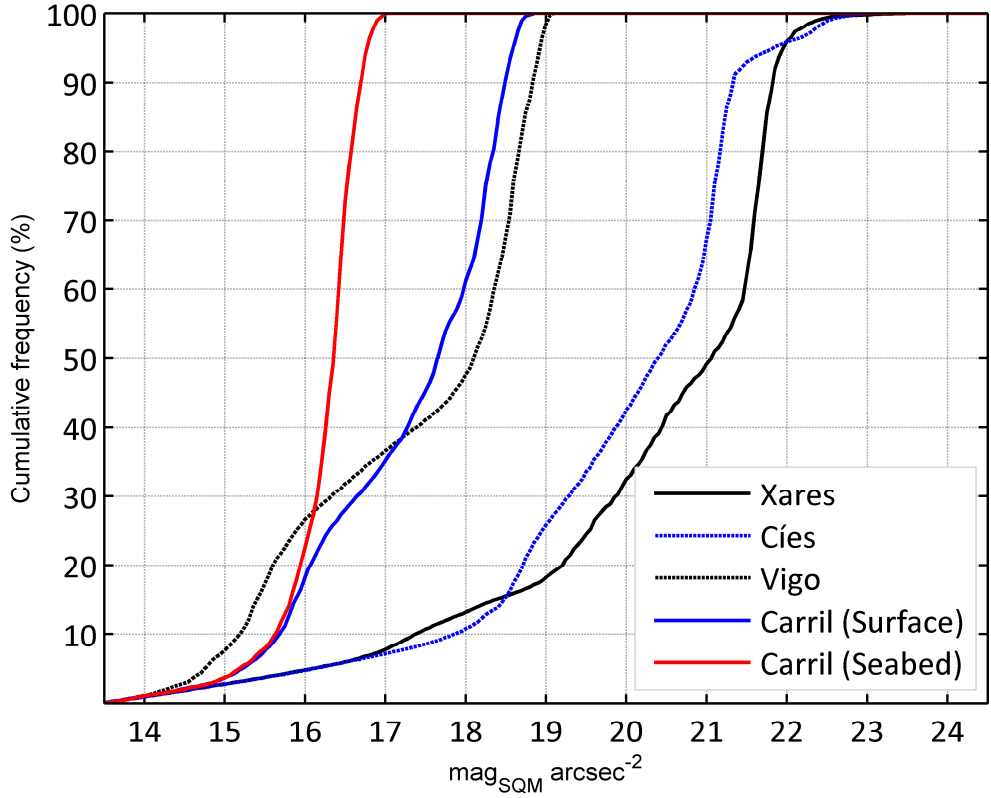

**Fig. 6**. Cumulative sums of the histograms shown in Fig. 5.

*3.2 NSB density plots*

Additional insights on the behaviour of the NSB at the different sites can be gained if the data are plotted with increasing degrees of temporal resolution. As a first step, Fig. 7 shows the NSB density plots (Pushnig et al, 2014a, 2014b; Bará, 2016; Posch et al., 2018). Each individual column of the plots corresponds to the histogram of the NSB data acquired at a given time of the night during the whole campaign period, displayed with time resolution 10 min and NSB resolution 0.05 $mag_{SQM}/arcsec^2$. The colorbars indicate the absolute number of records in each time and NSB interval. The series of almost vertical lines at the extremes of the plots correspond to the rapid variation of the zenithal sky brightness at dusk and dawn. These lines are located at different hours, reflecting the progressive decrease of daytime across the measurement period (July-August). In the density plots can also be seen the plateaus corresponding to the NSB values at the central hours of the night in clear and moonless nights, the secondary peaks due to the clouds, located about 3 $mag_{SQM}/arcsec^2$ below the plateaus (see particularly Vigo and Carril surface data), the intricate lines corresponding to the moonshine along the Moon phase cycle, and even some episodes of thick fog that reduced the measured sky brightness beyond its natural clear sky values (see e.g. points above 20 $mag_{SQM}/arcsec^2$ in the Illas Cíes sensor, Fig. 7, upper right). Another significant feature is the progressive darkening of the night sky in the Vigo metropolitan area along the first half of the night (Fig. 7, center), due to the gradual switch off of ornamental, commercial and residential lights, as well as to the reduction of vehicle traffic. Notice that neither the Carril surface data (otherwise similar to the Vigo ones) nor the Carril seabed ones, show signs of progressive darkening along the night, since the most important contribution to the recorded brightness is the constant radiance of the surveillance lampposts.

*3.3 NSB daily traces*

With an even greater time detail, the NSB daily traces or SQMgram plots (Fig. 8) show the daily evolution of the NSB at the measurement sites (Pushnig et al, 2014a, 2014b; Bará, 2016; Posch et al., 2018). Each column of these plots corresponds to a successive period of 24 hours, with the time increasing downwards in 10 min bins, starting from 12:00 UTC of a given day and ending at 11:59 UTC of the following one (i.e., the center of each column corresponds to midnight, 00:00 UTC). The colorbar units are $mag_{SQM}/arcsec^2$ and the value displayed at each 10 min bin corresponds to the average of the individual readings taking during that time interval at a rate of one every three minutes. The increase in the duration of the night along the campaign period is clearly visible. The slanted, broad, and lighter band present in the Xares and Illas Cíes plots reveals the changing contribution of moonshine. Note that both in the Vigo metropolitan area and in the Carril zone the Moon signal is not distinguishable, due to the overwhelming contribution of the artificial lights. Moonshine is completely obliterated at the aquaculture farm area by the intense and constant light flux of the surveillance luminaires.

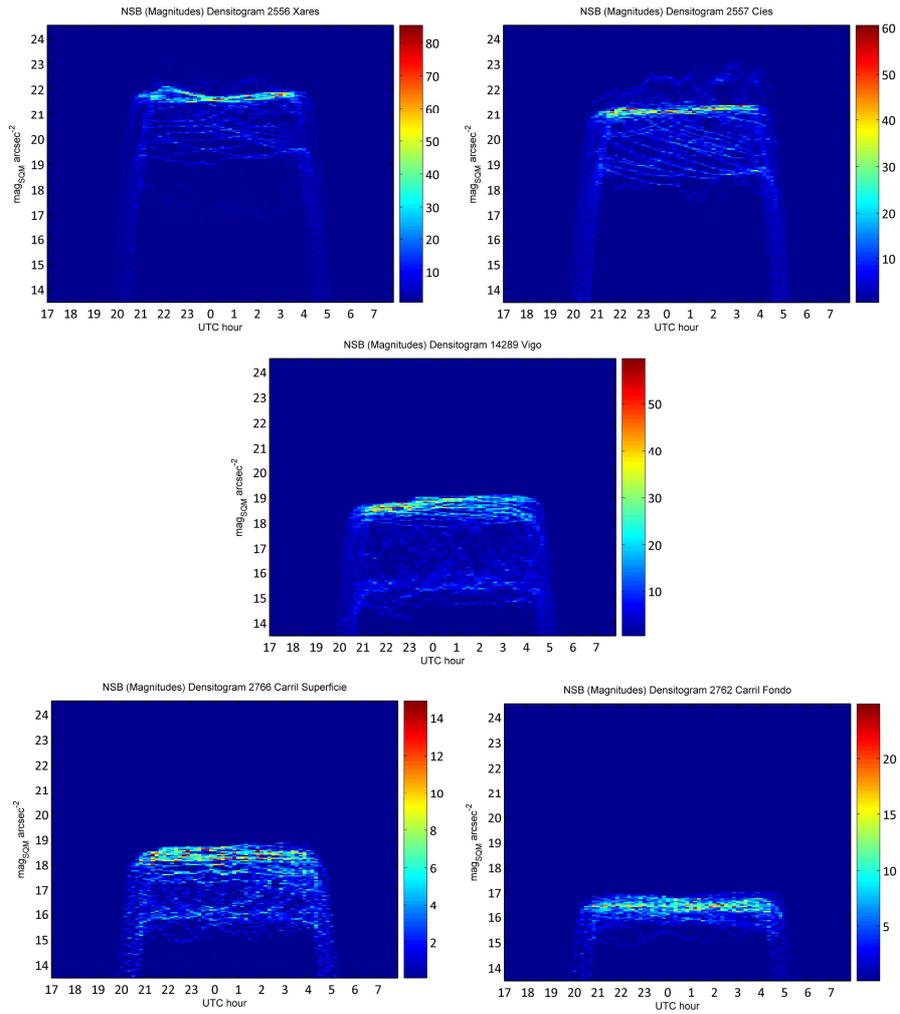

**Fig. 7**. Density plots of the recorded zenithal NSB values at the stations of Xares (upper left), Illas Cíes (upper right), Vigo (center), Carril surface (lower left), and Carril seabed (lower right). The time resolution is 10 min and the NSB resolution is 0.05 mag$_{SQM}$/arcsec$^2$. Colorbars indicate the absolute number of records in each time and NSB interval.

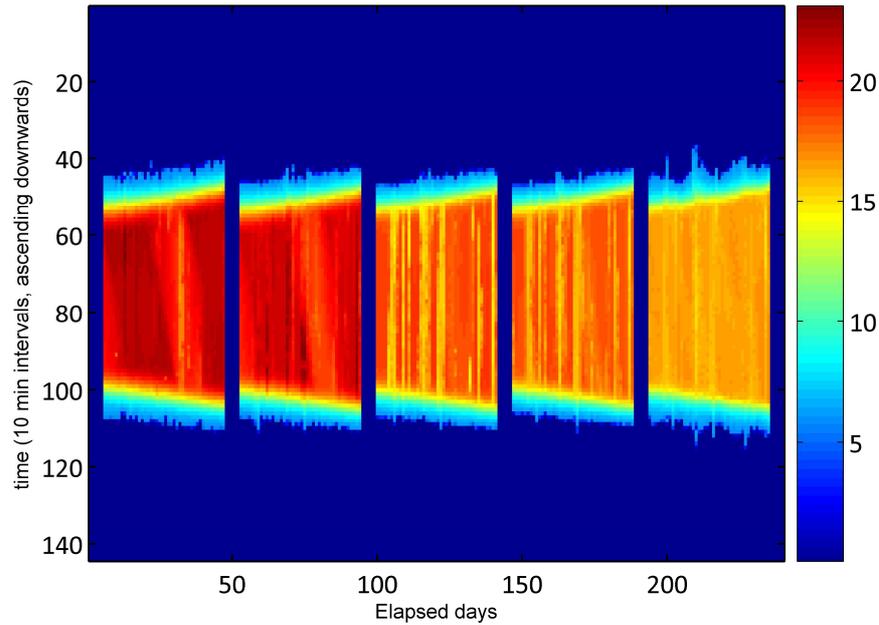

**Fig. 8**. (From left to right) Daily traces of the recorded zenithal night sky brightness values at the stations of Xares, Illas Cíes, Vigo, Carril surface, and Carril seabed, respectively. Colorbar units are $mag_{SQM}/arcsec^2$ and the value displayed at each 10 min bin corresponds to the average of the individual readings taking during that time interval at a rate of one every three minutes.

*3.4 Underwater NSB vs tide height*

The NSB recorded by the underwater sensor under different tide heights is displayed in Fig. 9. In this figure we only include readings taken in conditions of moonless astronomical night. To that end, the complete dataset was filtered to select the points corresponding to the times when the Sun and the Moon altitudes with respect to the horizon were below −18° and −5°, respectively. In Fig. 9 tide heights are expressed in meters above the lowest astronomical tide (LAT) baseline. In this reference frame the seabed sensor is located at +1.3 m height.

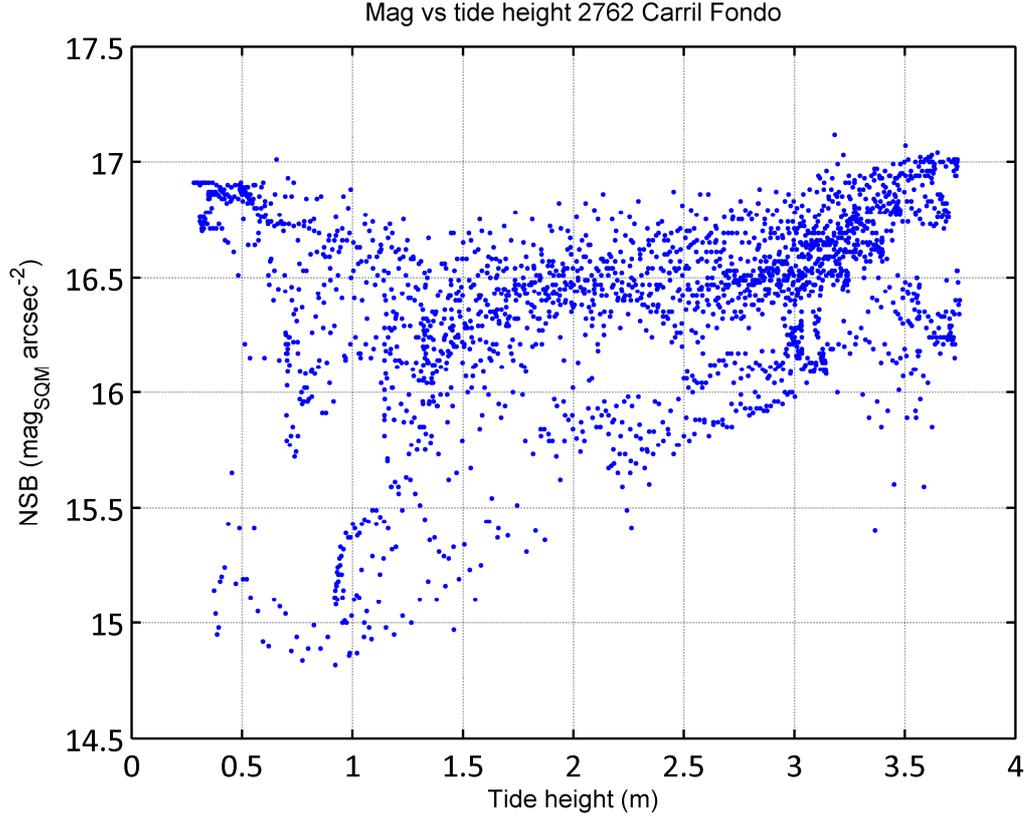

**Fig. 9**. NSB recorded by the underwater detector vs tide height above the lowest astronomical tide (LAT). The seabed detector is located at 1.3 m above LAT.

The recorded $mag_{SQM}$/arcsec$^2$ (a negative logarithmic expression of the NSB) tends to increase linearly with the height $h$ of the water column above the detector (in our case $h$ = tide height above LAT − 1.3 m), as it could be expected from the larger attenuation of the light beams after travelling longer paths in a strongly absorbing and scattering medium such as seawater. The downwelling irradiance diffuse attenuation coefficient, $K_d$, can be coarsely estimated from the fact that the NSB changes by about +0.6 $mag_{SQM}$/arcsec$^2$ when the tide height increases from 1.5 to 3.5 m. This corresponds to $K_d$=0.2763 m$^{-1}$, a value compatible with Jerlov's coastal water types 3C-5C for the wavelength range 500-550 nm (Solonenko and Mobley, 2015, Fig. 1). For a proper assessment of this result it is convenient to bear in mind that the $K_d$ definition involves the use of downwelling irradiances, whereas the SQM-LU-DL measurements are angularly weighted downwelling radiances. The attenuation with depth of both magnitudes, however, is expected to follow the same general trend.

## 4. Discussion

In this work we addressed the loss of the natural darkness of the night in a coastal area of rich biodiversity and relevant aquaculture farming activity, due to the artificial emissions of light, namely those of high luminance LED sources permanently switched on for surveillance purposes. Our results correspond to the measurements carried out in a two-month campaign in the summer of 2017.

As a proxy variable for characterizing the loss of darkness we used the zenithal night sky brightness measured with dedicated high-dynamic range detectors (SQM-LU-DL). Since the artificial night sky brightness is a highly variable parameter, because it is determined by the time course of the anthropogenic light emissions, the changing state of the atmosphere and, in case of an underwater detector, the height of the water column above the sensor, the concentration and type of suspended matter, and the characteristics of the surface sea waves, extended time datasets are necessary to achieve a full characterization of the status of any given site. A statistical parameter indicative of the degree of brightness of the zenithal night sky is the significant magnitude $m_{1/3}$ (Bará, 2016), that is, the mean value of the upper third of $mag_{SQM}/arcsec^2$ records taken with the Sun below below −18° (astronomical night) and the Moon below −5°, regardless of the Moon phase. For our measurement period the resulting $m_{1/3}$ significant magnitudes (in $mag_{SQM}/arcsec^2$) were 22.02 (Xares), 21.74 (Cíes), 18.92 (Vigo), 18.53 (Carril surface), and 16.77 (Carril seabed). Note that, according to Eq. 1, this means that the zenithal brightness corresponding to the $m_{1/3}$ significant magnitude at the Carril seabed is 125 times larger than in Xares, a nearly pristine natural site.

In a recent article, Holker et al. (2010a) have identified an emerging problem of great importance and urgency, namely light pollution as a diversity threat. Throughout evolution, most organisms have developed molecular circadian clocks controlled by natural day-night cycles. These clocks play key roles in growth, metabolism, reproduction, behaviour or predation. It is interesting to note that a large proportion of global biodiversity is considered nocturnal (>60% of all invertebrates). Light pollution threatens biodiversity through changing night habits. To our knowledge, no studies on how the lighting system in the Carril study area affects biodiversity have been carried out. There is also no specific data on the effect of artificial lighting on the production of bivalve mollusks in the parks of Carril. But there is evidence that the sustainable cultivation of the aforementioned species can be seriously altered by light pollution. In the last two years, cockle production has declined dramatically (Xunta de Galicia, 2017), partly attributable to a higher incidence of parasitic diseases and partly attributable to the drop in natural recruitment. It is interesting to mention that larvae of many species of bivalves are photonegative in their behaviour of settlement and metamorphosis (García-Lavandeira et al., 2005; Mesías-Gansbiller et al., 2013). In addition, the great influence of light on the reproductive process of some bivalve mollusks has recently been reported (Maneiro et al., 2016, 2017). The presence of continuous light and with different characteristics from natural light could be affecting the reproductive cycle of the mollusks grown in the parks and the success of the recruitment. On the other hand, the presence and activity of clam predators has increased significantly in the culture areas; coinciding with the installation of the night lighting system of the parks, a considerable increase in the predation of clams has been observed, mainly by gastropods, such as the species *Nassarius reticulatus* [Linnaeus 1758] (personal observation).

This study has several limitations that should be kept in mind. On the one hand, the photometric variable used herein to describe the anthropogenic loss of the natural night darkness is the spectral radiance integrated within the specific SQM spectral band, expressed in units of astronomical magnitudes per square arcsecond. Let us remind that the radiance (SI units $Wm^{-2}sr^{-1}$) is a characteristic function of the radiant field that describes the amount of radiant power per unit projected area and per unit solid angle centered around a given direction of propagation. In our case the SQM detectors were pointing at the zenith, hence the measured radiance corresponds to light propagating downwards within a Gaussian field of view of 20° FWHM. Many biological processes are strongly dependent on the radiance distribution, particularly those related to vision, orientation, navigation and displacements. Other relevant biological processes, however, depend on the surface density of radiant power (irradiance, SI units $Wm^{-2}$), which was not measured in our field campaign. The biological effects of light are also strongly dependent on the light spectrum; our measurements correspond to weighted integrals of the spectral radiance within the 400-650 nm SQM-LU-DL passband, and have no detailed information on the spectral composition of the light. However, the actual spectrum of the light arriving at the seabed can be easily estimated, based on the fact that it is produced by LED sources of known CCT, since the present solid state lighting technology based on phosphor-coated LED lamps produces standard, well characterized spectra, whose attenuation through different water types can be easily calculated (Solonenko and Mobley, 2015).

Another limitation is that our measurement campaign spanned several weeks in the summer of 2017, and not the whole year round. This fact, however, is somewhat less of a concern, since the luminance provided by the artificial sources is constant throughout the year, only varying the times of switching on and off, according to the seasonal variation of daylength. Although limited in time span, the results presented here are representative of the darkness disruption experienced all year round, since artificial lamps lit all night long are the dominant light sources at the farm seabed, including the intertidal and the first meters of the subtidal zone. The contribution of the surveillance lights to the seabed radiance is additive with respect to the remaining light sources, both the natural and the artificial from the shoreline urban settlements. The seasonal behavior of the latter can be seen in Bará (2016) and Posch et al. (2018).

Light at night is nominally used in this aquaculture area for several purposes, including improved visual performance of farmers and (perceived) safety against thefts. It seems, however, that there is a large room for substantially improving the present floodlighting approach, reducing both unnecessary energy consumption and detrimental environmental impacts. On the one hand, the effective workload at nighttime in this kind of farms is relatively small during extended periods of the year, and does not require the use of intense and permanent light sources. On the other hand, and contrary to the common belief, the actual effectiveness of artificial light at night as a theft deterrent is far from being proven. The majority of published studies reveal a lack of significant correlation between lighting levels and reduction of different types of crime (Marchant 2004, 2010; Steinbach et al., 2015). Lighting, by itself, seems to be a scarcely relevant factor for improving public security in a variety of societal settings. Oddly enough, an excess of badly directed light can act just the opposite way, severely hindering the visual performance of night guards. Passive surveillance systems (e.g. thermal cameras) as well as definite actions addressing the societal roots of burglary could provide better results than the present floodlighting approach, at a significantly lower environmental cost.

## 5. Conclusions

Surveillance lighting systems installed in shallow waters may contribute to a significant increase of the nighttime light levels throughout the whole water column, from the surface to the seabed. In the open field aquaculture area monitored in this study, the zenithal night brightness at the seabed reaches values 125 times larger than those expected for a terrestrial pristine natural site. This anthropogenic disruption of the natural night darkness has been shown to produce in related contexts unwanted, and presently uncontrolled effects, on important life processes across different taxa (metabolism, reproduction, foraging activity, predator-prey dynamics, orientation, etc). Given the potential severity of the environmental effects of artificial light at night, the dubious efficacy of light *per se* as a theft deterrence factor, and the possibility of using more efficient and environmentally friendly technologies for surveillance, the use of high-luminance lampposts permanently lit in coastal areas should be strongly discouraged and restricted as much as possible.

**Acknowledgments**

This work was partially supported by grant ED431B 2017/64, Xunta de Galicia/FEDER (S.B.), and by the 5th Edition of the IACOBUS programme, Agrupación Europea de Cooperación Territorial Galicia - Norte de Portugal (R.C.L.). The Galician Night Sky Brightness Measurement Network is a joint cooperation action of MeteoGalicia and Universidade de Santiago de Compostela (USC). Thanks are given to MeteoGalicia for providing the tide heigh data at the observation site, and to Manuel García-Turnes for kindly sharing the image in Fig.2. CITEUC is funded by National Funds through FCT - Foundation for Science and Technology (project: UID/Multi/00611/2013) and FEDER - European Regional Development Fund through COMPETE 2020 – Operational Programme Competitiveness and Internationalization (project: POCI-01-0145-FEDER-006922). We gratefully acknowledge the help and support of Agrupación de Parquistas de Carril (Carril Growers' Association), in particular José Luis Villanueva and Juan Vidal, to carry out the field work in the Carril farm site. This measurement campaign was developed in the framework of the USC - Wireless Galicia S.L. joint cooperation agreement "Light pollution monitoring in the metropolitan area of Vigo and in rural and coastal regions of Galicia".